\documentclass[superscriptaddress,twocolumn,floatfix]{revtex4}

\usepackage{graphicx}
\usepackage{amsmath}
\usepackage{color}

\begin{document}

\title{Metadevice for intensity modulation with sub-wavelength spatial resolution}

\author{Pablo Cencillo-Abad}
\affiliation{Optoelectronics Research Centre and Centre for Photonic Metamaterials, University of Southampton, Highfield, Southampton, SO17 1BJ, UK}


\author{Nikolay I. Zheludev}
\homepage{www.nanophotonics.org.uk}
\affiliation{Optoelectronics Research Centre and Centre for Photonic Metamaterials, University of Southampton, Highfield, Southampton, SO17 1BJ, UK}
\affiliation{Centre for Disruptive Photonic Technologies, School of Physical and Mathematical Sciences and The Photonics Institute, Nanyang Technological University, Singapore 637371}

\author{Eric Plum}
\email{erp@orc.soton.ac.uk}
\affiliation{Optoelectronics Research Centre and Centre for Photonic Metamaterials, University of Southampton, Highfield, Southampton, SO17 1BJ, UK}


\maketitle

\textbf{
Effectively continuous control over propagation of a beam of light requires light modulation with pixelation that is smaller than the optical wavelength.
Here we propose a spatial intensity modulator with sub-wavelength resolution in one dimension. The metadevice combines recent advances in reconfigurable nanomembrane metamaterials and coherent all-optical control of metasurfaces. It uses nanomechanical actuation of metasurface absorber strips placed near a mirror in order to control their interaction with light from perfect absorption to negligible loss, promising a path towards dynamic beam diffraction, light focusing and holography without unwanted diffraction artefacts.
}

\begin{figure}[h]
\includegraphics[width=85mm]{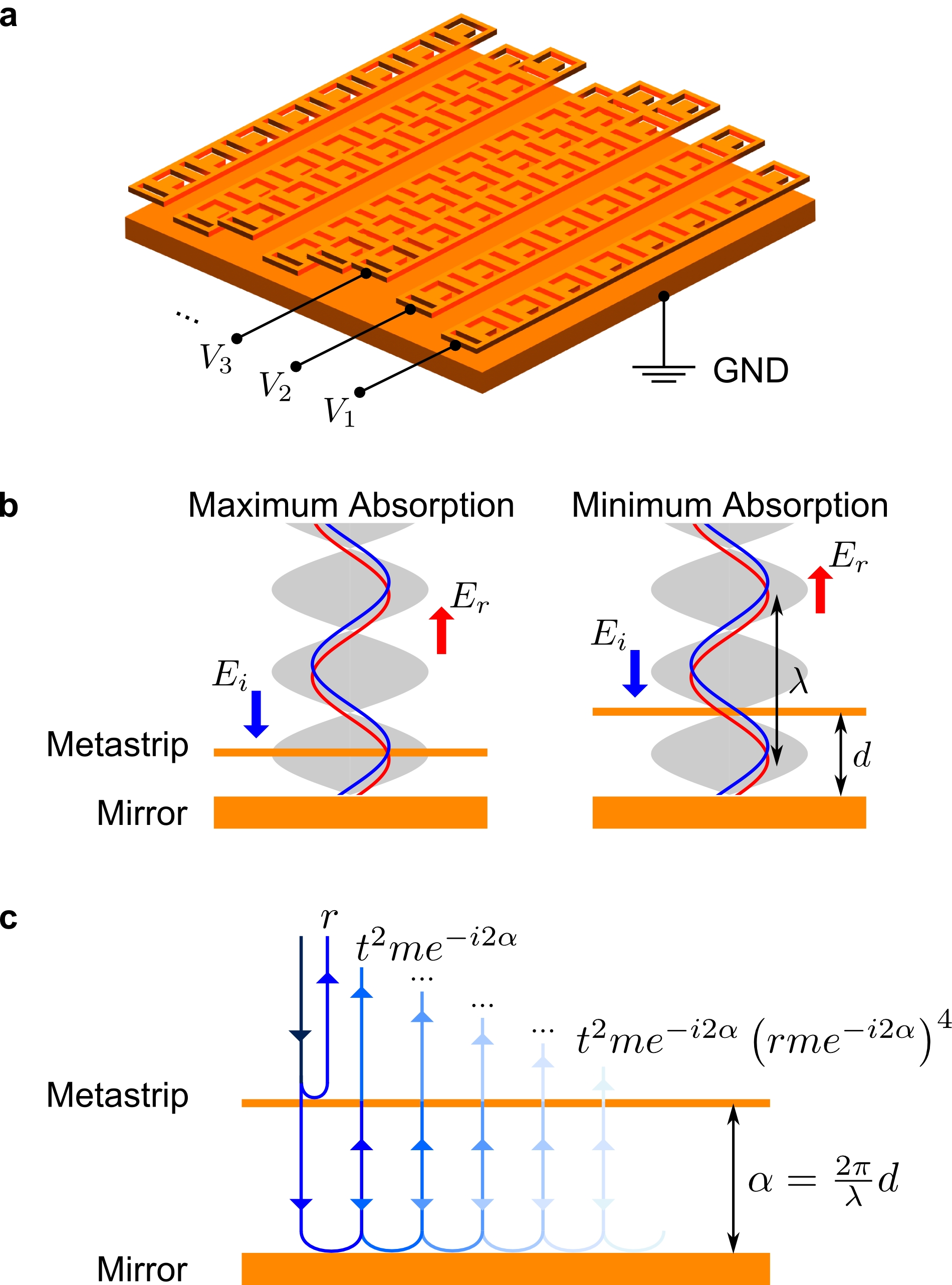}
\caption{\label{FigConcept}
\textbf{Intensity modulation based on coherent metasurface absorption.} (a)~Artistic impression of a spatial intensity modulator based on selective actuation of metastrips above a grounded metallic mirror (GND), where $V_i$ indicate metastrip actuation voltages. (b)~Incident $E_i$ and reflected $E_r$ electromagnetic waves form a standing wave (grey). The largest absorption occurs for metastrips placed at an electric field anti-node, while negligible absorption occurs for metastrips positioned at an electric field node. (c)~Multiple reflections at a metasurface with reflection coefficient $r$ and transmission coefficient $t$ placed at a distance $d$ in front of a mirror with reflection coefficient $m$.
}
\end{figure}

Spatial control over the intensity of light is the basis of optical components such as diffraction gratings, Fresnel zone plates and amplitude holograms. Dynamic spatial control over light intensity can --- in principle --- provide their functionalities on demand, however, established spatial light modulators based on liquid crystal or digital micromirror technology suffer from low resolution and unwanted diffraction due to their pixelation on the order of 10~$\mu$m \cite{efron_spatial_1994, lazarev_lcos_2012, yu_recent_2015}. Sub-wavelength resolution would be required to avoid unwanted diffraction and to achieve truly arbitrary and effectively continuous spatial control over the intensity of light. The required sub-wavelength scale structuring is a defining characteristic of metamaterials and metasurfaces and leads to locally homogeneous optical properties. Static spatial variation of metamaterial structures has given rise to the fields of transformation optics \cite{TransformationOptics_Science2006, TO_Cloak_Science2006, TO_Cloak_NatPhot_2007, TO_Cloak_NatMater_2009} and gradient metasurfaces \cite{Yu2011, SL2012, Walther2012, Roy2013, Ni2013, Lin2014, Chen2014a, Veksler2015}, while emerging technologies based on phase transitions \cite{driscoll_memory_2009, samson_metamaterial_2010}, nonlinearities \cite{cho_ultrafast_2009, dani_subpicosecond_2009}, 
coherent light-matter interactions \cite{zhang_controlling_2012, fang_ultrafast_2014, shi_coherent_2014, papaioannou_two-dimensional_2015} and nanomechanical actuation \cite{lapine_structural_2009, tao_reconfigurable_2009, pryce_highly_2010, zhu_switchable_2011, ou_reconfigurable_2011, ou_electromechanically_2013, valente_reconfiguring_2015, Bruce_NL_RPM_2016, valente_magneto-electro-optical_2015, Padilla_coherent_IR_RPM_2013}
now enable the temporal modulation of metamaterial properties \cite{zheludev_obtaining_2015}.

Here we propose a metadevice that controls coherent absorption of light by selective actuation of metamaterial nanostructures to provide arbitrary dynamic control over its reflectivity with sub-wavelength spatial resolution in one dimension, see Fig.~\ref{FigConcept}a.

Absorbtion of light in a planar absorber of sub-wavelength thickness is limited to 50\%, if it is illuminated from only one side \cite{MetasurfaceMaxAbsorption_2012}. However, if it is placed within a standing wave formed by counterpropagating coherent waves, then absorption will depend on its position relative to the standing wave's nodes and anti-nodes. At an electric field node, the planar structure cannot interact with the wave and therefore absorption is nominally zero, while absorption at an electric field anti-node can in principle reach 100\% (ref.~\onlinecite{zhang_controlling_2012}). Projection of images onto a metasurface absorber with coherent light has enabled all-optical control of light absorption with diffraction-limited resolution \cite{papaioannou_two-dimensional_2015}, however, this approach cannot beat the diffraction limit and it requires a stable interferometer as well as coherent light.
Instead, we propose a mechanically reconfigurable nanoscale version of a Salisbury screen \cite{SalisburyScreen_1988, Munk_2000_FSS, SalisburyScreen_Engheta_2002, MagMirror2007}, where individually addressable absorbing metamaterial strips are placed at a variable nanoscale distance from a mirror in order to exploit coherent perfect absorption and transparency on demand. As illustrated by Fig.~\ref{FigConcept}, the incident light and multiple reflections by mirror and metasurface form a standing wave, without a macroscopic interferometer and even for incoherent illumination provided that the coherence length is large compared to the metasurface-to-mirror spacing $d$ (measured from the middle of the metasurface to the mirror surface). Furthermore, the spatial resolution of the metadevice will be determined by the pitch of the metamaterial strip actuators, i.e.~it will be controlled by nanofabrication technology, rather than the diffraction limit. Thermal, electric, magnetic and optical actuation of such metamaterial strips \cite{ou_reconfigurable_2011, ou_electromechanically_2013, valente_reconfiguring_2015, Bruce_NL_RPM_2016} 
as well as actuation of similar optomechanical nanostructures \cite{thijssen_plasmon_2013, thijssen_parallel_2014, yamaguchi_electrically_2014, thijssen_plasmomechanical_2015} has been reported.
Recently, selective actuation of individual metamaterial strips has been demonstrated \cite{Pablo_NanoMeta_2015} and suggested as a method for spatial phase modulation \cite{PhaseRPM_numerical_2016}.
Structures for the optical part of the spectrum are typically based on a dielectric membrane of nanoscale thickness (e.g.~silicon nitride) which is coated by a plasmonic material (e.g.~gold) or a high index dielectric (e.g.~silicon), which is then structured by reactive ion etching and focused ion beam milling to create the metamaterial pattern and strip actuators with or without the original dielectric layer \cite{zheludev_reconfigurable_2016}.

\section*{Results and Discussion}

\begin{figure}[h]
\includegraphics[width=85mm]{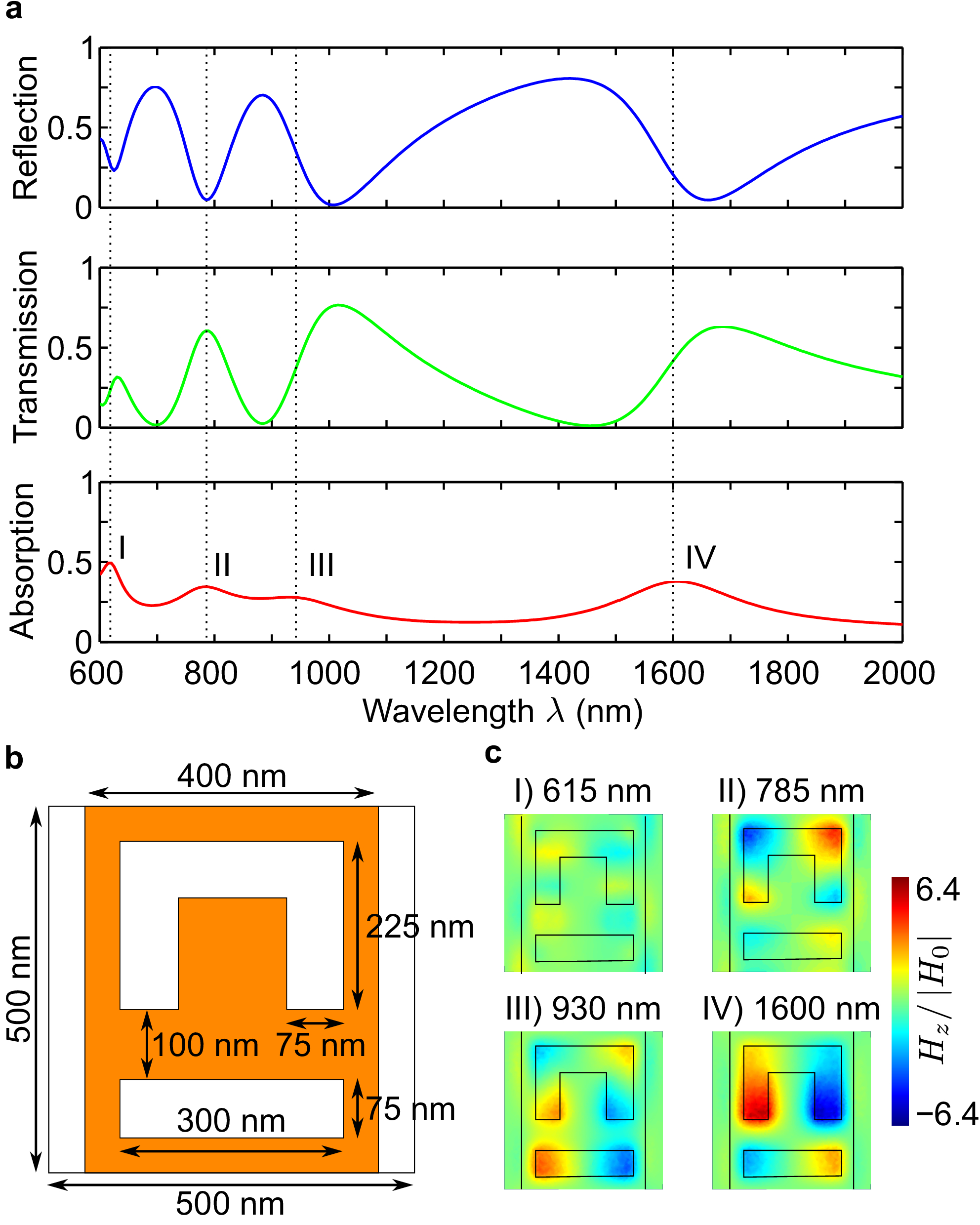}
\caption{\label{FigASR}
\textbf{The metamaterial and its resonances.} (a)~Reflection, transmission and absorption of the flat metasurface without mirror. (b)~Schematic of the unit cell. (c)~Modes of excitation at the absorption maxima in terms of the instantaneous magnetic field $H_z$ 10~nm above the gold surface relative to the incident wave's magnetic field amplitude $|H_0|$.
}
\end{figure}

\begin{figure*}[h]
\includegraphics[width=165mm]{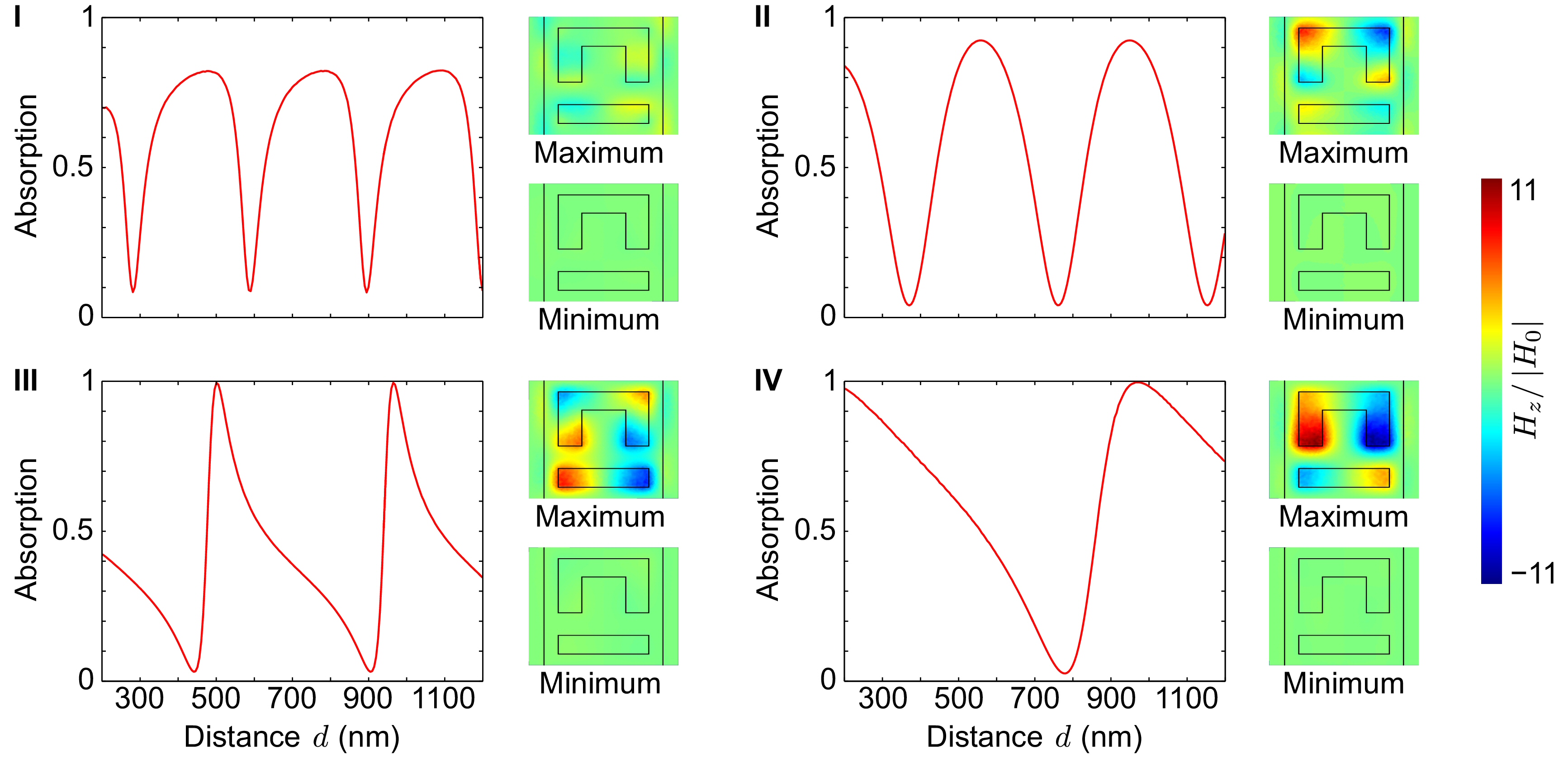}
\caption{\label{FigSpacing}
\textbf{Dependence of absorption on the metasurface-to-mirror spacing $d$} at wavelengths of (I) 615~nm, (II) 785~nm, (III) 930~nm and (IV) 1600~nm. Colour maps show the instantaneous magnetic field $H_z$ 10~nm above the gold surface relative to the incident wave's magnetic field amplitude $|H_0|$ for absorption maxima and minima.
}
\end{figure*}

\begin{figure*}[h]
\includegraphics[width=165mm]{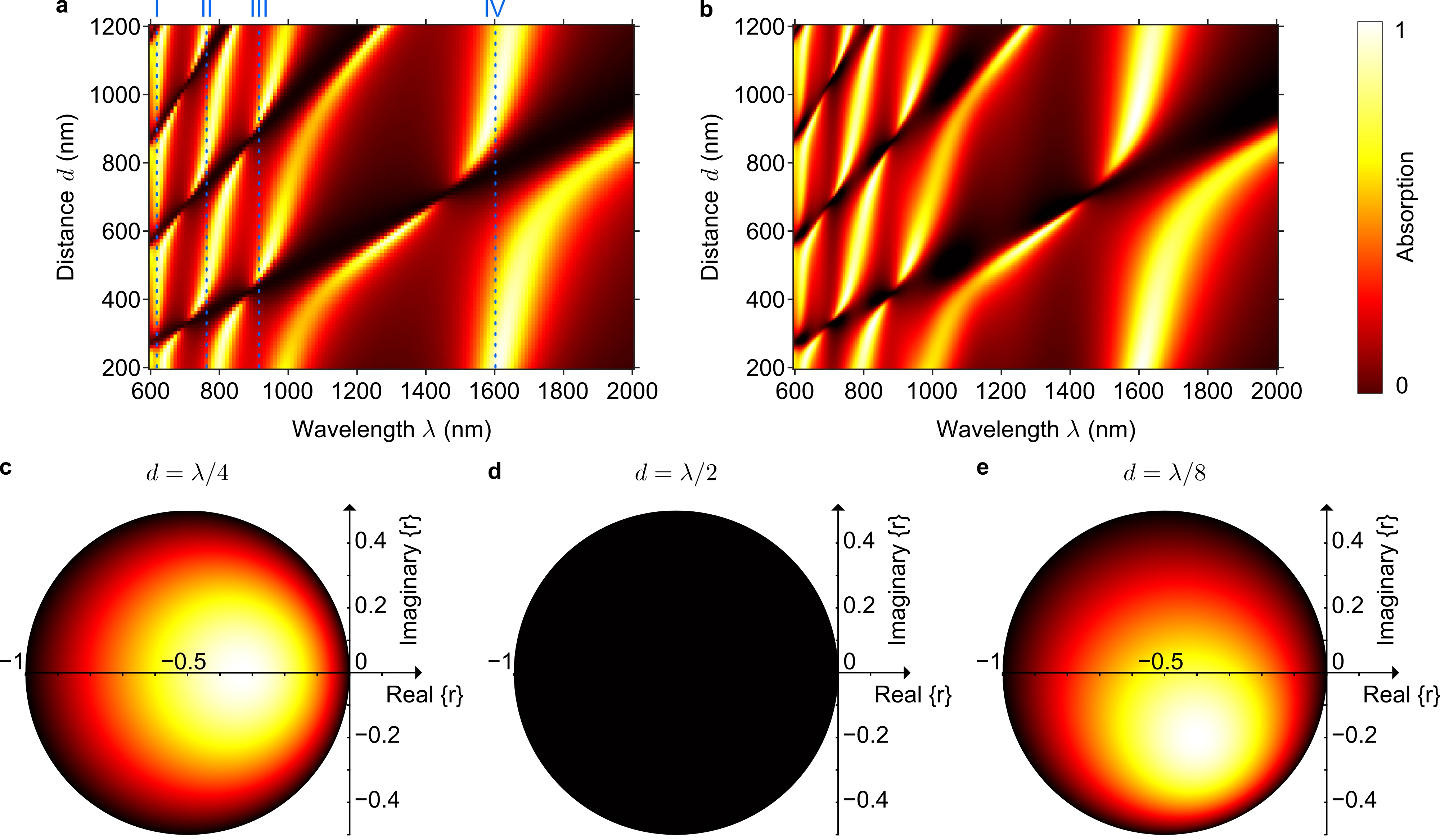}
\caption{\label{FigMaps}
\textbf{Dependence of absorption on metasurface-to-mirror spacing $d$ and wavelength $\lambda$} based on (a) numerical modelling of the entire metadevice and (b) the semi-analytical model given by equation (\ref{Eq_full_absorption}). (c-e) Dependence of absorption on the metasurface scattering coefficient $r$ for metasurface-to-mirror distances (c) $d=\lambda/4$, (d) $\lambda/2$ and (e) an intermediate spacing of $\lambda/8$ based on the analytical model given by equation (\ref{Eq_simplified_absorption}).
}
\end{figure*}

The metadevice considered here consists of a reconfigurable metasurface in front of a metallic mirror, see Fig.~\ref{FigConcept}a. The metasurface is composed of parallel gold strips of 400~nm width and 50~nm thickness that are separated by 100~nm gaps and perforated with asymmetrically split ring apertures, see Fig.~\ref{FigASR}. A unit cell size of 500~nm was chosen to enable non-diffracting operation in the red to infrared spectral range where gold is highly reflective. Asymmetrically split ring apertures were chosen as they are well-studied \cite{combinatorial_ASR} and known to provide coherent perfect absorption \cite{zhang_controlling_2012}. The size of the split rings was determined by the available space on the gold strips prescribing a minimum feature size of 50 nm that can be routinely achieved with both focused ion beam milling and electron beam lithography. The optical properties of metasurface and metadevice were simulated for normal incidence illumination by a coherent plane wave using finite element modelling (COMSOL Multiphysics 4.4) in three dimensions, approximating the device with metastrips that have prescribed displacements and infinite length. We consider linearly polarised light with the electric field parallel to the strips in all cases to avoid the excitation of metal-insulator-metal waveguide modes. The electric permittivity of gold is computed from a Drude-Lorentz model with 3 oscillators \cite{liu_plasmonic_2008}.
Fig.~\ref{FigASR}a illustrates the optical properties of the flat metasurface 
without the backing mirror. The split ring aperture array has rich transmission, reflection and absorption spectra with several resonances corresponding to the fundamental and higher order modes of the split ring slits, see Fig.~\ref{FigASR}c.

The optical properties of the structure change dramatically when it is combined with a mirror. For an optically thick mirror, transmission is zero and reflectivity $R$ and absorption $A$ are given by the superposition of the waves that are multiply reflected by mirror and metasurface as illustrated by Fig.~\ref{FigConcept}c. Neglecting near-field effects,
\begin{equation}
A=1-R=1-\left|r+\frac{e^{-i2\alpha}mt^2}{1-e^{-i2\alpha}mr}\right|^2 \label{Eq_full_absorption}
\end{equation}
where $r$ and $t$ are the complex Fresnel reflection and transmission coefficients of the flat metasurface, $m$ is the reflection coefficient of the mirror and $\alpha=2\pi d/\lambda$ is the phase accumulated during propagation of the wave of wavelength $\lambda$ from metasurface to mirror.

Assuming an ideal mirror ($m=-1$), the metadevice absorption is that of the metasurface. In this case, the electric field at the metasurface position --- which controls absorption --- corresponds to the superposition of the incident wave with the wave that is transmitted by the metasurface and then multiply reflected between mirror and metasurface. The metasurface absorption $A$ is proportional to the square of the resulting electric field enhancement.
For an ideal planar metasurface ($t=r+1$),
\begin{equation}
A= \frac{1-\cos{2\alpha}}{\left|1+e^{-i2\alpha}r\right|^2} 2 A_0 \label{Eq_simplified_absorption}
\end{equation}
where $A_0=1-|r|^2-|r+1|^2$ is the metasurface absorption without the mirror.
Notably, this analytical model predicts that absorption vanishes completely due to electric field cancellation at the metasurface when $\alpha$ is a multiple of $\pi$, i.e.~when the metasurface-to-mirror spacing $d$ becomes a multiple of $\lambda/2$.

As illustrated by Fig.~\ref{FigSpacing}, this behaviour is confirmed by numerical simulations. Absorption of the metadevice is strongly dependent on the metasurface-to-mirror spacing and reduces to few \% for wavelengths corresponding approximately to multiples of $\lambda/2$. The small residual reflectivity and 10s-of-nm deviations in the spectral position of the absorption minima result from (i) the metasurface not being perfectly planar due to having a finite thickness of 50~nm ($t\simeq r+1$) and (ii) the gold mirror not being an ideal mirror due to small absorption losses and slight deviations from reflection with a $\pi$ phase change ($m\simeq -1$). For intermediate spacings, the metasurface's resonant modes are excited (compare Figs.~\ref{FigASR}c and \ref{FigSpacing}) and the strength of the metasurface excitation is controlled by the distance $d$ that determines the interference of incident and (multiply) reflected waves on the metasurface. Constructive interference results in stronger excitation and higher levels of absorption ($>99\%$ for selected wavelengths) than for the metasurface without the backing mirror, compare with Fig.~\ref{FigASR}a. Destructive interference suppresses metasurface excitation resulting in negligible absorption as discussed above. We note that absorption as a function of the metasurface-to-mirror spacing is generally periodic with period $\lambda/2$ (provided that $d$ is small compared to the coherence length of the illuminating light), however, some deviations due to near-field interactions between metasurface and mirror are observed for small gaps $d$ approaching 200~nm.

Fig.~\ref{FigMaps}a,b shows the absorption spectrum of the metadevice as a function of the metasurface-to-mirror spacing $d$, where panel (a) shows results for numerical modelling of the entire metadevice, while panel (b) shows semi-analytical results based on equation (\ref{Eq_full_absorption}), where the metasurface properties, $r$ and $t$, were determined separately by numerical modelling and the reflectivity of the gold mirror $m$ was calculated using the Fresnel equations. The results are almost indistinguishable and they reveal that the metadevice's absorption can be strongly modulated in wide spectral bands around the metasurface's resonances I-IV (see Fig.~\ref{FigSpacing}). These bands of operation are separated by the metasurface's transmission minima at 700, 885 and 1455~nm (compare to Fig.~\ref{FigASR}a) and the absorption wavelength can be moved continuously across each band by adjusting the metasurface-to-mirror spacing, resulting in a wavelength-tuneable metadevice. Notably, the bands of operation include the red part of the visible spectrum and the main telecommunications bands around 1310~nm and 1550~nm wavelength.
The absence of absorption when the metasurface-to-mirror spacing coincides with a multiple of $\lambda/2$ corresponds to the straight dark bands across the colormaps.

Fig.~\ref{FigMaps}c-e illustrates based on equation (\ref{Eq_simplified_absorption}) how the metadevice absorption depends on the metasurface's complex scattering coefficient $r$ for characteristic metasurface-to-mirror spacings $d$. As should be expected, absorption vanishes for lossless metasurfaces, which satisfy $|r+0.5|=0.5$, and, as shown above, absorption vanishes for $d=\lambda/2$ regardless of the metasurface's scattering coefficient. For lossy metasurfaces, the dependence of absorption on $d$ is symmetric for real scattering coefficients (as in Fig.~\ref{FigSpacing} I, II) and asymmetric for complex scattering coefficients (as in Fig.~\ref{FigSpacing} III, IV).

\begin{figure*}[h]
\includegraphics[width=165mm]{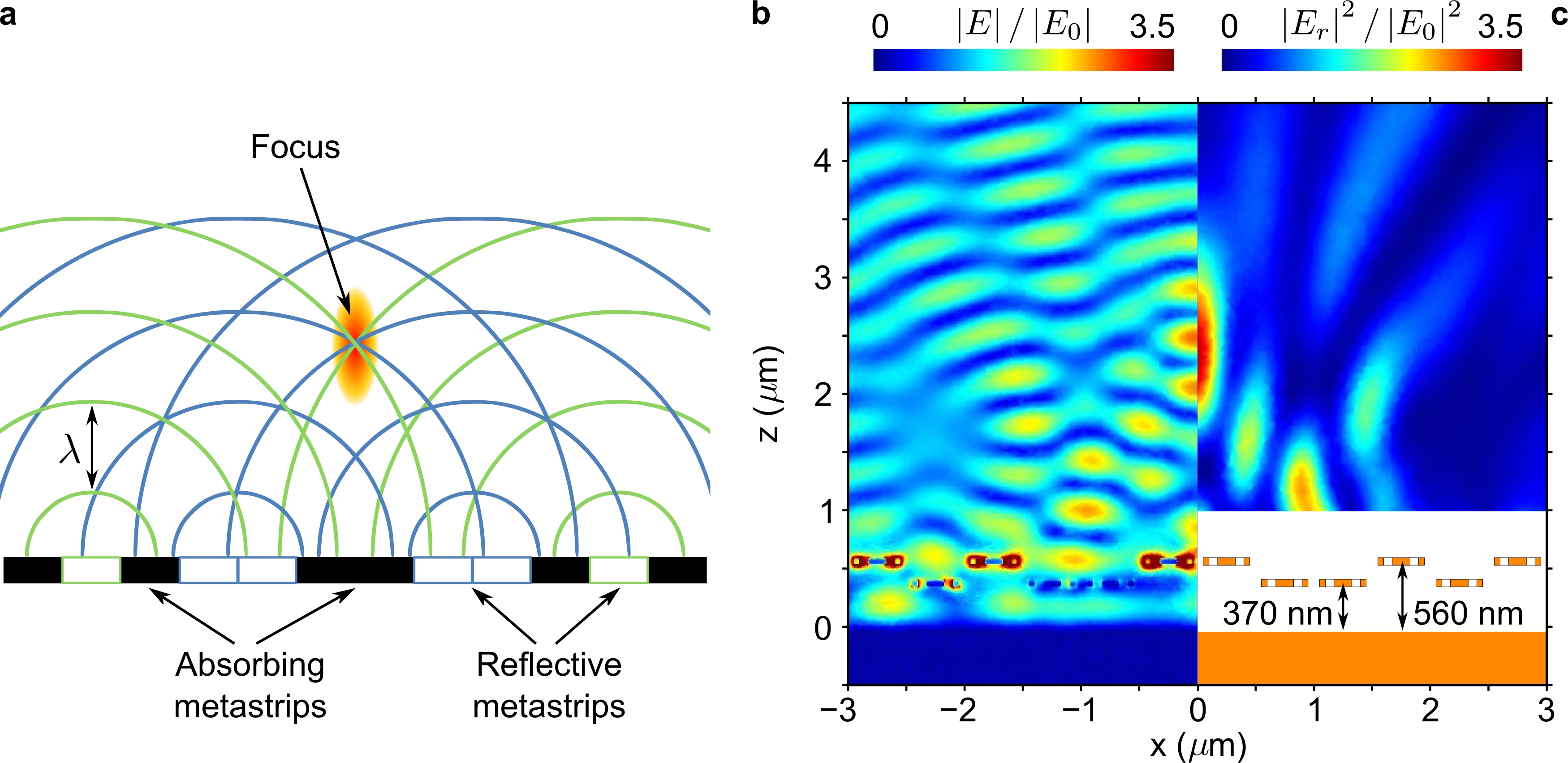}
\caption{\label{FigFresnelZonePlate}
\textbf{Fresnel zone plate application} of a metadevice with 12 metastrips operating at $\lambda=785$~nm wavelength. (a)~Basic operating principle of the zone plate: all reflective areas are sources of scattered fields, which interfere constructively at the focal spot. (b,c)~A focal spot at a distance of $2~\mu$m results from metastrip-to-mirror spacings of $d=370$~nm (negligible absorption) and 560~nm (perfect absorption) as shown. The colour maps show (b)~the total electric field amplitude $|E|$ and (c)~the square of the reflected electric field's amplitude $|E_r|^2$ normalised to the incident wave $E_0$.
}
\end{figure*}

As indicated on Fig.~\ref{FigConcept}a, actuation of individual metamaterial strips can be driven by electrostatic forces resulting from grounding the mirror and application of an electric potential to an individual nanowire as in commercial grating light valves \cite{bloom_grating_1997, solgaard_photonic_2009}. In-plane strip deformation can be neglected as the strips deform about 100$\times$ more easily towards the mirror than towards their neighbours due to their perforated center and their width/thickness aspect ratio of 8 (ref.~\onlinecite{MechanicsBook2009}). 
Spatial intensity modulation results from positioning different metastrips at different distances from the mirror. This is illustrated by Fig.~\ref{FigFresnelZonePlate} for a metastrip configuration that corresponds to a Fresnel zone plate operating at $\lambda$=785~nm wavelength. The zone plate consists of reflective and absorbing areas arranged in such a way that all scattered fields will constructively interfere at the intended focus as illustrated by panel (a). Here, the zone plate is realised by placing the metastrips at positions of either negligible ($d=370$~nm) or almost complete ($d=560$~nm) absorption as shown. In order to achieve a line focus 2~$\mu$m above the metadevice, light scattered by the reflective areas must constructively interfere on the focal line. Therefore, all strips for which the strip-to-focus optical path length $p_i$ satisfies $0 \leq p_i/\lambda - N_i - C < 0.5$, where $N_i$ is an integer and $C=0.6$ is an arbitrarily chosen real number, were set to the reflective position, while all other strips were set to the absorbing position.
The simulations show that coupling between neighboring strips is sufficiently small to allow weak excitation of strips at $d=370$~nm and strong excitation of strips at $d=560$~nm even in the most extreme case when these reflective and absorbing strips alternate, see panel (b). The same panel also shows the complex pattern of standing waves that forms above the actuated metadevice due to interference of incident and scattered fields.
Panel (c) shows the focal spot that forms 2~$\mu$m above the metadevice in its Fresnel zone plate configuration due to constructive interference of the fields scattered by the weakly absorbing strips.

Regarding the experimental feasibility of the metadevice, actuation of metastrips of similar complexity and characteristic dimensions has already been demonstrated \cite{zheludev_reconfigurable_2016}. Metastrips will generally be fixed at both ends and therefore voltage application will result in metastrip bending towards the ground plane, where only the middle section of each strip will achieve the desired displacement. The size of this optically useful middle section of the metastrips scales with the overall strip length and it can be enlarged by making the end sections of the strips more elastic (e.g. thinner). Permanent metastrip deformation was not found to be a major issue in prior work and could be addressed by adding an elastic dielectric supporting layer (e.g. silicon or silicon nitride). However, we note that electrostatic forces will overcome the elastic restoring force of the metastrips when the metastrip-to-mirror spacing is reduced to approximately half of its original size \cite{ou_electromechanically_2013}. In order to achieve reliable electrostatic operation without such accidental switching of the device, metastrips should only be displaced by significantly less than half of their original distance from the ground plane. Slight differences in response and rest position of different metastrips could arise from fabrication imperfections, but could be offset using adjusted actuation voltages after device calibration. Achievable modulation rates of metastrips are limited by their mechanical resonances and the associated resonance frequencies scale inversely proportional to the square of the metastrip length. Typical metastrips of 10s of $\mu$m length have mechanical resonances at 100s of kHz to MHz and could have actuation voltages of several V if placed 100s of nm away from a ground plane \cite{ou_electromechanically_2013, valente_magneto-electro-optical_2015, zheludev_reconfigurable_2016}.

\section*{Conclusion}

In summary, we demonstrate that nanoscale actuation of metamaterial strips placed at a nanoscale distance from a mirror could be the basis for a spatial light modulator with sub-wavelength spatial resolution in one dimension and --- in principle --- unlimited optical contrast resulting from complete reflection and coherent perfect absorption. We argue that recent breakthroughs in reconfigurable nanomembrane metamaterials and coherent all-optical control of metasurfaces make such a device a realistic proposition. Potential application areas of such metadevices include dynamic diffraction, focusing and attenuation of light as well as holography.


\begin{thebibliography}{10}
\expandafter\ifx\csname url\endcsname\relax
  \def\url#1{\texttt{#1}}\fi
\expandafter\ifx\csname urlprefix\endcsname\relax\def\urlprefix{URL }\fi
\providecommand{\bibinfo}[2]{#2}
\providecommand{\eprint}[2][]{\url{#2}}

\bibitem{efron_spatial_1994}
\bibinfo{editor}{Efron, U.} (ed.) \emph{\bibinfo{title}{Spatial Light Modulator
  Technology: Materials, Devices, and Applications}} (\bibinfo{publisher}{CRC
  Press}, \bibinfo{year}{1994}).

\bibitem{lazarev_lcos_2012}
\bibinfo{author}{Lazarev, G.}, \bibinfo{author}{Hermerschmidt, A.},
  \bibinfo{author}{Kr{\"u}ger, S.} \& \bibinfo{author}{Osten, S.}
\newblock \emph{\bibinfo{title}{Optical Imaging and Metrology: Advanced
  Technologies}}, chap. \bibinfo{chapter}{1. LCOS Spatial Light Modulators:
  Trends and Applications}, \bibinfo{pages}{1--29}
  (\bibinfo{publisher}{Wiley-VCH Verlag GmbH \& Co. KGaA},
  \bibinfo{year}{2012}).

\bibitem{yu_recent_2015}
\bibinfo{author}{Yu, H.} \emph{et~al.}
\newblock \bibinfo{title}{Recent advances in wavefront shaping techniques for
  biomedical applications}.
\newblock \emph{\bibinfo{journal}{Curr. Appl. Phys.}}
  \textbf{\bibinfo{volume}{15}}, \bibinfo{pages}{632--641}
  (\bibinfo{year}{2015}).

\bibitem{TransformationOptics_Science2006}
\bibinfo{author}{Leonhardt, U.}
\newblock \bibinfo{title}{Optical conformal mapping}.
\newblock \emph{\bibinfo{journal}{Science}} \textbf{\bibinfo{volume}{312}},
  \bibinfo{pages}{1777--1780} (\bibinfo{year}{2006}).

\bibitem{TO_Cloak_Science2006}
\bibinfo{author}{Schurig, D.} \emph{et~al.}
\newblock \bibinfo{title}{Metamaterial electromagnetic cloak at microwave
  frequencies}.
\newblock \emph{\bibinfo{journal}{Science}} \textbf{\bibinfo{volume}{314}},
  \bibinfo{pages}{977--980} (\bibinfo{year}{2006}).

\bibitem{TO_Cloak_NatPhot_2007}
\bibinfo{author}{Cai, W.}, \bibinfo{author}{Chettiar, U.~K.},
  \bibinfo{author}{Kildishev, A.~V.} \& \bibinfo{author}{Shalaev, V.~M.}
\newblock \bibinfo{title}{Optical cloaking with metamaterials}.
\newblock \emph{\bibinfo{journal}{Nat. Photonics}}
  \textbf{\bibinfo{volume}{1}}, \bibinfo{pages}{224--227}
  (\bibinfo{year}{2007}).

\bibitem{TO_Cloak_NatMater_2009}
\bibinfo{author}{Valentine, J.}, \bibinfo{author}{Li, J.},
  \bibinfo{author}{Zentgraf, T.}, \bibinfo{author}{Bartal, G.} \&
  \bibinfo{author}{Zhang, X.}
\newblock \bibinfo{title}{An optical cloak made of dielectrics}.
\newblock \emph{\bibinfo{journal}{Nat. Mater.}} \textbf{\bibinfo{volume}{8}},
  \bibinfo{pages}{568--571} (\bibinfo{year}{2009}).

\bibitem{Yu2011}
\bibinfo{author}{Yu, N.~F.} \emph{et~al.}
\newblock \bibinfo{title}{Light propagation with phase discontinuities:
  generalized laws of reflection and refraction}.
\newblock \emph{\bibinfo{journal}{Science}} \textbf{\bibinfo{volume}{334}},
  \bibinfo{pages}{333--337} (\bibinfo{year}{2011}).

\bibitem{SL2012}
\bibinfo{author}{Sun, S.~L.} \emph{et~al.}
\newblock \bibinfo{title}{High-efficiency broadband anomalous reflection by
  gradient meta-surfaces}.
\newblock \emph{\bibinfo{journal}{Nano Lett.}} \textbf{\bibinfo{volume}{12}},
  \bibinfo{pages}{6223--6229} (\bibinfo{year}{2012}).

\bibitem{Walther2012}
\bibinfo{author}{Walther, B.}, \bibinfo{author}{Helgert, C.},
  \bibinfo{author}{Rockstuhl, C.}, \bibinfo{author}{Setzpfandt, F.} \&
  \bibinfo{author}{Eilenberger, F.}
\newblock \bibinfo{title}{Spatial and spectral light shaping with
  metamaterials}.
\newblock \emph{\bibinfo{journal}{Adv. Mater.}} \textbf{\bibinfo{volume}{24}},
  \bibinfo{pages}{6300--6304} (\bibinfo{year}{2012}).

\bibitem{Roy2013}
\bibinfo{author}{Roy, T.}, \bibinfo{author}{Nikolaenko, A.~E.} \&
  \bibinfo{author}{Rogers, E.~T.}
\newblock \bibinfo{title}{A meta-diffraction-grating for visible light}.
\newblock \emph{\bibinfo{journal}{J. Opt.}} \textbf{\bibinfo{volume}{15}},
  \bibinfo{pages}{85101} (\bibinfo{year}{2013}).

\bibitem{Ni2013}
\bibinfo{author}{Ni, X.~J.}, \bibinfo{author}{Ishii, S.},
  \bibinfo{author}{Kildishev, A.~V.} \& \bibinfo{author}{Shalaev, V.~M.}
\newblock \bibinfo{title}{Ultra-thin, planar, babinet-inverted plasmonic
  metalenses}.
\newblock \emph{\bibinfo{journal}{Light. Sci. Appl.}}
  \textbf{\bibinfo{volume}{2}}, \bibinfo{pages}{e72} (\bibinfo{year}{2013}).

\bibitem{Lin2014}
\bibinfo{author}{Lin, D.~M.}, \bibinfo{author}{Fan, P.~Y.},
  \bibinfo{author}{Hasman, E.} \& \bibinfo{author}{Brongersma, M.~L.}
\newblock \bibinfo{title}{Dielectric gradient metasurface optical elements}.
\newblock \emph{\bibinfo{journal}{Science}} \textbf{\bibinfo{volume}{345}},
  \bibinfo{pages}{298--302} (\bibinfo{year}{2014}).

\bibitem{Chen2014a}
\bibinfo{author}{Chen, W.~T.} \emph{et~al.}
\newblock \bibinfo{title}{High-efficiency broadband meta-hologram with
  polarization-controlled dual images}.
\newblock \emph{\bibinfo{journal}{Nano Lett.}} \textbf{\bibinfo{volume}{14}},
  \bibinfo{pages}{225--230} (\bibinfo{year}{2014}).

\bibitem{Veksler2015}
\bibinfo{author}{Veksler, D.}, \bibinfo{author}{Maguid, E.},
  \bibinfo{author}{Shitrit, N.}, \bibinfo{author}{Ozeri, D.} \&
  \bibinfo{author}{Kleiner, V.}
\newblock \bibinfo{title}{Multiple wavefront shaping by metasurface based on
  mixed random antenna groups}.
\newblock \emph{\bibinfo{journal}{ACS Photonics}} \textbf{\bibinfo{volume}{2}},
  \bibinfo{pages}{661--667} (\bibinfo{year}{2015}).

\bibitem{driscoll_memory_2009}
\bibinfo{author}{Driscoll, T.} \emph{et~al.}
\newblock \bibinfo{title}{Memory metamaterials}.
\newblock \emph{\bibinfo{journal}{Science}} \textbf{\bibinfo{volume}{325}},
  \bibinfo{pages}{1518--1521} (\bibinfo{year}{2009}).

\bibitem{samson_metamaterial_2010}
\bibinfo{author}{S\'{a}mson, Z.~L.} \emph{et~al.}
\newblock \bibinfo{title}{Metamaterial electro-optic switch of nanoscale
  thickness}.
\newblock \emph{\bibinfo{journal}{Appl. Phys. Lett.}}
  \textbf{\bibinfo{volume}{96}}, \bibinfo{pages}{143105}
  (\bibinfo{year}{2010}).

\bibitem{cho_ultrafast_2009}
\bibinfo{author}{Cho, D.~J.} \emph{et~al.}
\newblock \bibinfo{title}{Ultrafast modulation of optical metamaterials}.
\newblock \emph{\bibinfo{journal}{Opt. Express}} \textbf{\bibinfo{volume}{17}},
  \bibinfo{pages}{17652--17657} (\bibinfo{year}{2009}).

\bibitem{dani_subpicosecond_2009}
\bibinfo{author}{Dani, K.~M.} \emph{et~al.}
\newblock \bibinfo{title}{Subpicosecond optical switching with a negative index
  metamaterial}.
\newblock \emph{\bibinfo{journal}{Nano Lett.}} \textbf{\bibinfo{volume}{9}},
  \bibinfo{pages}{3565--3569} (\bibinfo{year}{2009}).

\bibitem{zhang_controlling_2012}
\bibinfo{author}{Zhang, J.}, \bibinfo{author}{MacDonald, K.~F.} \&
  \bibinfo{author}{Zheludev, N.~I.}
\newblock \bibinfo{title}{Controlling light-with-light without nonlinearity}.
\newblock \emph{\bibinfo{journal}{Light Sci. Appl.}}
  \textbf{\bibinfo{volume}{1}}, \bibinfo{pages}{e18} (\bibinfo{year}{2012}).

\bibitem{fang_ultrafast_2014}
\bibinfo{author}{Fang, X.} \emph{et~al.}
\newblock \bibinfo{title}{Ultrafast all-optical switching via coherent
  modulation of metamaterial absorption}.
\newblock \emph{\bibinfo{journal}{Appl. Phys. Lett.}}
  \textbf{\bibinfo{volume}{104}}, \bibinfo{pages}{141102}
  (\bibinfo{year}{2014}).

\bibitem{shi_coherent_2014}
\bibinfo{author}{Shi, J.} \emph{et~al.}
\newblock \bibinfo{title}{Coherent control of snell's law at metasurfaces}.
\newblock \emph{\bibinfo{journal}{Opt. Express}} \textbf{\bibinfo{volume}{22}},
  \bibinfo{pages}{21051--21060} (\bibinfo{year}{2014}).

\bibitem{papaioannou_two-dimensional_2015}
\bibinfo{author}{Papaioannou, M.}, \bibinfo{author}{Plum, E.},
  \bibinfo{author}{Valente, J.}, \bibinfo{author}{Rogers, E.} \&
  \bibinfo{author}{Zheludev, N.}
\newblock \bibinfo{title}{Two-dimensional control of light with light on
  metasurfaces}.
\newblock \emph{\bibinfo{journal}{Light Sci. Appl.}}
  \textbf{\bibinfo{volume}{5}}, \bibinfo{pages}{e16070} (\bibinfo{year}{2015}).

\bibitem{lapine_structural_2009}
\bibinfo{author}{Lapine, M.} \emph{et~al.}
\newblock \bibinfo{title}{Structural tunability in metamaterials}.
\newblock \emph{\bibinfo{journal}{Appl. Phys. Lett.}}
  \textbf{\bibinfo{volume}{95}}, \bibinfo{pages}{084105}
  (\bibinfo{year}{2009}).

\bibitem{tao_reconfigurable_2009}
\bibinfo{author}{Tao, H.} \emph{et~al.}
\newblock \bibinfo{title}{Reconfigurable terahertz metamaterials}.
\newblock \emph{\bibinfo{journal}{Phys. Rev. Lett.}}
  \textbf{\bibinfo{volume}{103}}, \bibinfo{pages}{147401}
  (\bibinfo{year}{2009}).

\bibitem{pryce_highly_2010}
\bibinfo{author}{Pryce, I.~M.}, \bibinfo{author}{Aydin, K.},
  \bibinfo{author}{Kelaita, Y.~A.}, \bibinfo{author}{Briggs, R.~M.} \&
  \bibinfo{author}{Atwater, H.~A.}
\newblock \bibinfo{title}{Highly strained compliant optical metamaterials with
  large frequency tunability}.
\newblock \emph{\bibinfo{journal}{Nano Lett.}} \textbf{\bibinfo{volume}{10}},
  \bibinfo{pages}{4222--4227} (\bibinfo{year}{2010}).

\bibitem{zhu_switchable_2011}
\bibinfo{author}{Zhu, W.~M.} \emph{et~al.}
\newblock \bibinfo{title}{Switchable magnetic metamaterials using
  micromachining processes}.
\newblock \emph{\bibinfo{journal}{Adv. Mater.}} \textbf{\bibinfo{volume}{23}},
  \bibinfo{pages}{1792--1796} (\bibinfo{year}{2011}).

\bibitem{ou_reconfigurable_2011}
\bibinfo{author}{Ou, J.~Y.}, \bibinfo{author}{Plum, E.},
  \bibinfo{author}{Jiang, L.} \& \bibinfo{author}{Zheludev, N.~I.}
\newblock \bibinfo{title}{Reconfigurable photonic metamaterials}.
\newblock \emph{\bibinfo{journal}{Nano Lett.}} \textbf{\bibinfo{volume}{11}},
  \bibinfo{pages}{2142--2144} (\bibinfo{year}{2011}).

\bibitem{ou_electromechanically_2013}
\bibinfo{author}{Ou, J.~Y.}, \bibinfo{author}{Plum, E.},
  \bibinfo{author}{Zhang, J.} \& \bibinfo{author}{Zheludev, N.~I.}
\newblock \bibinfo{title}{An electromechanically reconfigurable plasmonic
  metamaterial operating in the near-infrared}.
\newblock \emph{\bibinfo{journal}{Nat. Nanotechnol.}}
  \textbf{\bibinfo{volume}{8}}, \bibinfo{pages}{252--255}
  (\bibinfo{year}{2013}).

\bibitem{valente_reconfiguring_2015}
\bibinfo{author}{Valente, J.}, \bibinfo{author}{Ou, J.~Y.},
  \bibinfo{author}{Plum, E.}, \bibinfo{author}{Youngs, I.~J.} \&
  \bibinfo{author}{Zheludev, N.~I.}
\newblock \bibinfo{title}{Reconfiguring photonic metamaterials with currents
  and magnetic fields}.
\newblock \emph{\bibinfo{journal}{Appl. Phys. Lett.}}
  \textbf{\bibinfo{volume}{106}}, \bibinfo{pages}{111905}
  (\bibinfo{year}{2015}).

\bibitem{Bruce_NL_RPM_2016}
\bibinfo{author}{Ou, J.~Y.}, \bibinfo{author}{Plum, E.},
  \bibinfo{author}{Zhang, J.} \& \bibinfo{author}{Zheludev, N.~I.}
\newblock \bibinfo{title}{Giant nonlinearity of an optically reconfigurable
  plasmonic metamaterial}.
\newblock \emph{\bibinfo{journal}{Adv. Mater.}} \textbf{\bibinfo{volume}{28}},
  \bibinfo{pages}{729--733} (\bibinfo{year}{2016}).

\bibitem{valente_magneto-electro-optical_2015}
\bibinfo{author}{Valente, J.}, \bibinfo{author}{Ou, J.~Y.},
  \bibinfo{author}{Plum, E.}, \bibinfo{author}{Youngs, I.~J.} \&
  \bibinfo{author}{Zheludev, N.~I.}
\newblock \bibinfo{title}{A magneto-electro-optical effect in a plasmonic nanowire material}.
\newblock \emph{\bibinfo{journal}{Nat. Commun.}}
  \textbf{\bibinfo{volume}{6}}, \bibinfo{pages}{7021}
  (\bibinfo{year}{2015}).

\bibitem{Padilla_coherent_IR_RPM_2013}
\bibinfo{author}{Liu, X.} \& \bibinfo{author}{Padilla, W.~J.}
\newblock \bibinfo{title}{Dynamic manipulation of infrared radiation with mems
  metamaterials}.
\newblock \emph{\bibinfo{journal}{Adv. Opt. Mater.}}
  \textbf{\bibinfo{volume}{1}}, \bibinfo{pages}{559--562}
  (\bibinfo{year}{2013}).

\bibitem{zheludev_obtaining_2015}
\bibinfo{author}{Zheludev, N.~I.}
\newblock \bibinfo{title}{Obtaining optical properties on demand}.
\newblock \emph{\bibinfo{journal}{Science}} \textbf{\bibinfo{volume}{348}},
  \bibinfo{pages}{973--974} (\bibinfo{year}{2015}).

\bibitem{MetasurfaceMaxAbsorption_2012}
\bibinfo{author}{Thongrattanasiri, S.}, \bibinfo{author}{Koppens, F. H.~L.} \&
  \bibinfo{author}{de~Abajo, F. J.~G.}
\newblock \bibinfo{title}{Complete optical absorption in periodically patterned
  graphene}.
\newblock \emph{\bibinfo{journal}{Phys. Rev. Lett.}}
  \textbf{\bibinfo{volume}{108}}, \bibinfo{pages}{047401}
  (\bibinfo{year}{2012}).

\bibitem{SalisburyScreen_1988}
\bibinfo{author}{Fante, R.~L.} \& \bibinfo{author}{McCormack, M.~T.}
\newblock \bibinfo{title}{Reflection properties of the salisbury screen}.
\newblock \emph{\bibinfo{journal}{IEEE Trans. Antennas. Propag.}}
  \textbf{\bibinfo{volume}{36}}, \bibinfo{pages}{1443--1454}
  (\bibinfo{year}{1988}).

\bibitem{Munk_2000_FSS}
\bibinfo{author}{Munk, B.~A.}
\newblock \emph{\bibinfo{title}{Frequency Selective Surfaces: Theory and
  Design}} (\bibinfo{publisher}{John Wiley \& Sons}, \bibinfo{year}{2000}).

\bibitem{SalisburyScreen_Engheta_2002}
\bibinfo{author}{Engheta, N.}
\newblock \bibinfo{title}{Thin absorbing screens using metamaterial surfaces}.
\newblock In \emph{\bibinfo{booktitle}{Antennas and Propagation Society
  International Symposium}}, vol.~\bibinfo{volume}{2},
  \bibinfo{pages}{392--395} (\bibinfo{year}{2002}).

\bibitem{MagMirror2007}
\bibinfo{author}{Schwanecke, A.~S.}, \bibinfo{author}{Fedotov, V.~A.},
  \bibinfo{author}{V.~V.~Khardikov, Y.~C., S. L.~Prosvirnin} \&
  \bibinfo{author}{Zheludev, N.~I.}
\newblock \bibinfo{title}{Optical magnetic mirrors}.
\newblock \emph{\bibinfo{journal}{J. Opt. A: Pure Appl. Opt.}}
  \textbf{\bibinfo{volume}{9}}, \bibinfo{pages}{L1--L2} (\bibinfo{year}{2007}).

\bibitem{thijssen_plasmon_2013}
\bibinfo{author}{Thijssen, R.}, \bibinfo{author}{Verhagen, E.},
  \bibinfo{author}{Kippenberg, T.~J.} \& \bibinfo{author}{Polman, A.}
\newblock \bibinfo{title}{Plasmon nanomechanical coupling for nanoscale
  transduction}.
\newblock \emph{\bibinfo{journal}{Nano Lett.}} \textbf{\bibinfo{volume}{13}},
  \bibinfo{pages}{3293--3297} (\bibinfo{year}{2013}).

\bibitem{thijssen_parallel_2014}
\bibinfo{author}{Thijssen, R.}, \bibinfo{author}{Kippenberg, T.~J.},
  \bibinfo{author}{Polman, A.} \& \bibinfo{author}{Verhagen, E.}
\newblock \bibinfo{title}{Parallel transduction of nanomechanical motion using
  plasmonic resonators}.
\newblock \emph{\bibinfo{journal}{ACS Photonics}} \textbf{\bibinfo{volume}{1}},
  \bibinfo{pages}{1181--1188} (\bibinfo{year}{2014}).

\bibitem{yamaguchi_electrically_2014}
\bibinfo{author}{Yamaguchi, K.}, \bibinfo{author}{Fujii, M.},
  \bibinfo{author}{Okamoto, T.} \& \bibinfo{author}{Haraguchi, M.}
\newblock \bibinfo{title}{Electrically driven plasmon chip: Active plasmon
  filter}.
\newblock \emph{\bibinfo{journal}{Appl. Phys. Express}}
  \textbf{\bibinfo{volume}{7}}, \bibinfo{pages}{012201} (\bibinfo{year}{2014}).

\bibitem{thijssen_plasmomechanical_2015}
\bibinfo{author}{Thijssen, R.}, \bibinfo{author}{Kippenberg, T.~J.},
  \bibinfo{author}{Polman, A.} \& \bibinfo{author}{Verhagen, E.}
\newblock \bibinfo{title}{Plasmomechanical resonators based on dimer
  nanoantennas}.
\newblock \emph{\bibinfo{journal}{Nano Lett.}} \textbf{\bibinfo{volume}{15}},
  \bibinfo{pages}{3971--3976} (\bibinfo{year}{2015}).

\bibitem{Pablo_NanoMeta_2015}
\bibinfo{author}{Cencillo-Abad, P.}, \bibinfo{author}{Ou, J.~Y.},
  \bibinfo{author}{Valente, J.}, \bibinfo{author}{Plum, E.} \&
  \bibinfo{author}{Zheludev, N.~I.}
\newblock \bibinfo{title}{Randomly addressable reconfigurable photonic
  metamaterials.}
\newblock In \emph{\bibinfo{booktitle}{5th Int. Topical Meeting on
  Nanophotonics and Metamaterials, Seefeld in Tirol, Austria}}
  (\bibinfo{year}{2015}).

\bibitem{PhaseRPM_numerical_2016}
\bibinfo{author}{Cencillo-Abad, P.}, \bibinfo{author}{Plum, E.},
  \bibinfo{author}{Rogers, E. T.~F.} \& \bibinfo{author}{Zheludev, N.~I.}
\newblock \bibinfo{title}{Spatial optical phase-modulating metadevice with
  subwavelength pixelation}.
\newblock \emph{\bibinfo{journal}{Opt. Express}} \bibinfo{pages}{in press}
  (\bibinfo{year}{2016}).

\bibitem{zheludev_reconfigurable_2016}
\bibinfo{author}{Zheludev, N.~I.} \& \bibinfo{author}{Plum, E.}
\newblock \bibinfo{title}{Reconfigurable nanomechanical photonic
  metamaterials}.
\newblock \emph{\bibinfo{journal}{Nat. Nanotechnol.}}
  \textbf{\bibinfo{volume}{11}}, \bibinfo{pages}{16--22}
  (\bibinfo{year}{2016}).

\bibitem{combinatorial_ASR}
\bibinfo{author}{Plum, E.} \emph{et~al.}
\newblock \bibinfo{title}{A combinatorial approach to metamaterials discovery}.
\newblock \emph{\bibinfo{journal}{J. Opt.}} \textbf{\bibinfo{volume}{13}},
  \bibinfo{pages}{055102} (\bibinfo{year}{2011}).

\bibitem{liu_plasmonic_2008}
\bibinfo{author}{Liu, Z.} \emph{et~al.}
\newblock \bibinfo{title}{Plasmonic nanoantenna arrays for the visible}.
\newblock \emph{\bibinfo{journal}{Metamaterials}} \textbf{\bibinfo{volume}{2}},
  \bibinfo{pages}{45--51} (\bibinfo{year}{2008}).

\bibitem{bloom_grating_1997}
\bibinfo{author}{Bloom, D.~M.}
\newblock \bibinfo{title}{Grating light valve: revolutionizing display
  technology}.
\newblock \emph{\bibinfo{journal}{Proc. SPIE}} \textbf{\bibinfo{volume}{3013}},
  \bibinfo{pages}{165--171} (\bibinfo{year}{1997}).

\bibitem{solgaard_photonic_2009}
\bibinfo{author}{Solgaard, O.}
\newblock \emph{\bibinfo{title}{Photonic Microsystems: Micro and Nanotechnology
  Applied to Optical Devices and Systems}} (\bibinfo{publisher}{Springer
  Science \& Business Media}, \bibinfo{year}{2009}).

\bibitem{MechanicsBook2009}
\bibinfo{author}{Timoshenko, S.~P.} \& \bibinfo{author}{Gere, J.~M.}
\newblock \emph{\bibinfo{title}{Theory of Elastic Stability}}
  (\bibinfo{publisher}{Dover Publications Inc.}, \bibinfo{year}{2009}).

\end{thebibliography}

\section*{Acknowledgements}

This work is supported by the MOE Singapore (grant MOE2011-T3-1-005) and the UK's Engineering and Physical Sciences Research Council (grants EP/G060363/1 and EP/M009122/1).
The authors acknowledge the use of the IRIDIS High Performance Computing Facility, and associated support services at the University of Southampton, in the completion of this work.
The data from this paper can be obtained from the University of Southampton ePrints research repository: http://dx.doi.org/10.5258/SOTON/398132

\section*{Author contributions}

P.C.A. caried out the simulations and performed the data analysis. P.C.A. and E.P. conceived the study and drafted the manuscript. E.P. and N.I.Z. supervised the work and all authors read and approved the manuscript.

\section*{Additional information}

\textbf{Competing financial interests:} The authors declare no
competing financial interests.

\end{document}